\documentclass[prd,preprint,tightenlines,superscriptaddress,showpacs,aps,amsmath,amssymb,nofootinbib]{revtex4}
\usepackage{graphicx} 
\usepackage{epsfig} 
\usepackage{hyperref}

\newcommand{\beq}{\begin{equation}}
\newcommand{\eeq}{\end{equation}}

\def\lsim{\raise0.3ex\hbox{$\;<$\kern-0.75em\raise-1.1ex\hbox{$\sim\;$}}}
\def\gsim{\raise0.3ex\hbox{$\;>$\kern-0.75em\raise-1.1ex\hbox{$\sim\;$}}}
\def\eps{\varepsilon}
\def\theta{\vartheta}

\begin{document}

\title{Extremely High Energy Neutrinos from Cosmic Strings } 

\author{Veniamin Berezinsky}
\email{venya.berezinsky@lngs.infn.it}
\affiliation{INFN, Laboratori Nazionali del Gran Sasso, I--67010 Assergi (AQ), Italy.}
\author{Eray Sabancilar}
\email{eray.sabancilar@tufts.edu}
\author{Alexander Vilenkin}
\email{vilenkin@cosmos.phy.tufts.edu}
\affiliation{Institute of Cosmology, Department of Physics and Astronomy, Tufts University, Medford, Massachusetts 02155, USA.}

\begin{abstract}
Superstring theory and other supersymmetric theories predict the
existence of relatively light, weakly interacting scalar particles,
called moduli,  with a universal form of coupling to matter.  
Such particles can be emitted from cusps of cosmic
strings, where extremely large Lorentz factors are achieved
momentarily. Highly boosted modulus bursts emanating from cusps
subsequently decay into gluons, they generate parton cascades 
which in turn produce large numbers of pions and then neutrinos. 
Due to very large Lorentz factors, 
extremely high energy neutrinos, up to the Planck scale and 
above, are produced.  For some model parameters, the predicted flux 
of neutrinos with energies $\gtrsim 10^{21}$~eV is observable by 
JEM-EUSO and by the future large radio detectors LOFAR and SKA. 
\end{abstract}

\maketitle


 \section{Introduction}
\label{sec:introduction}

Cosmic strings could be formed as topological defects in the early
universe. They are predicted in a wide class of particle physics
models and can
produce a variety of observational effects. These include 
gravitational lensing, linear discontinuities in the cosmic microwave
background, and gravitational radiation, both in the form of a
stochastic background and localized bursts. (For a review of cosmic
strings, see, e.g., \cite{Vilenkin-book,Copeland-Kibble}.) 

Strings predicted in many grand unified models respond to external
electromagnetic fields as thin superconducting wires \cite{Witten}.
As they move through cosmic magnetic fields, such strings develop
electric currents. Oscillating loops of current-carrying string emit
highly boosted charged particles from cusps -- short segments where
the string velocity momentarily gets very close to the speed of
light. The emitted particles and their decay products can then be observed
as high-energy cosmic rays \cite{BOSV} and gamma ray bursts
\cite{Paczynski,Berezinsky-Hnatyk-Vilenkin,Cheng-Yu-Harko}.   

A phenomenon closely related to string superconductivity is the
development of a bosonic condensate around the string core
\cite{Witten}.  For example, a condensate of Standard Model Higgs
particles could form around strings in some models.  The Higgses would
then be copiously produced at cusps, and their decay products could
reach the Earth as cosmic rays \cite{Vachaspati}.  

Here we shall discuss an alternative mechanism of cosmic ray
production, which does not assume string superconductivity or Higgs
condensates.  It relies on the existence of moduli -- relatively
light, weakly coupled scalar fields, predicted in supersymmetric
particle theories, including string theory. Moduli would be copiously
radiated by oscillating loops of string at early cosmic times, when
the loops are smaller than the modulus Compton wavelenth, 
$L < 1/m$, and their frequency of oscillation is greater
than the modulus mass. The emitted moduli 
may affect the big bang nucleosynthesis  as they decay into photons 
and baryons, contribute to dark matter and to diffuse gamma ray
background, resulting in stringent constraints  
on both the cosmic string tension and the
modulus mass, when moduli are assumed to have gravitational-strength 
couplings to matter 
\cite{Damour-Vilenkin-moduli,Peloso-Sorbo,Babichev,Sabancilar}.
However, the couplings may in fact be much stronger, 
in which case the constraints from moduli radiation may be
significantly relaxed \cite{Sabancilar}. 
Such strongly coupled moduli appear to be quite generic in string 
theory landscape 
\cite{large-volume2,Conlon-Quevedo,Frey-Maharana,Burgess,
Goldberger-Wise,Brummer}, and this case is of particular interest 
for production of Extremely High Energy (EHE) cosmic rays and
neutrinos.  

At later times, moduli can only be emitted from cusps, resulting in
sharp bursts of high-energy moduli. Eventually moduli decay into
standard model particles, and their decay products can be observed as
cosmic rays  with energies above $10^{21}$~eV. 

The great interest of cosmic strings, and more 
generically topological defects, to high energy neutrino astronomy 
is based on tremendous energies of neutrinos accessible for these 
sources. While astrophysical sources can accelerate particles to 
energies $10^{21} - 10^{22}$~eV at most, topological defects can
produce particles, including neutrinos, up to the Planck scale and above. 
Many observational methods of neutrino detection, in particular 
radio observations and observation of fluorescent light from 
space, are possible only above  $\sim 10^{20}$~eV.  Detection of such 
high energy neutrinos  
can by itself be considered as a signature of neutrinos from 
topological defects or other top-down scenarios. The production 
of EHE particles is a very generic property of topological 
defects, cosmic strings in particular, but large fluxes of such
particles are produced only in exceptional cases \cite{BOSV}, 
\cite{BBV} - \cite{BV-mirror}.       



In this paper, we shall treat the modulus mass and coupling constant
and the string tension as free parameters. We shall estimate the
EHE neutrino flux resulting from modulus decays
and indicate some values of the parameters that can yield observable
fluxes. The paper is organized as follows. In Section II, we review
modulus emission from cosmic string cusps (a more
detailed derivation is given in the Appendix). In Section III, we discuss 
modulus decay, EHE neutrino production, including beaming, and 
propagation in the universe. 
In Section IV, we review the size 
distribution of cosmic string loops and calculate the rate of bursts
and the diffuse flux of EHE
neutrinos. We also discuss here the upper bound on the neutrino flux, 
resulting from the diffuse gamma ray background observations.  
At the end of that section we give two illustrative examples of neutrino  
fluxes for different values of the model parameters.
Finally, we discuss EHE neutrino detection. Conclusions
are presented in Section V.   

  
\section{Modulus Radiation from Strings}
\label{ModulusRadiation}


The effective action for a modulus field $\phi$ interacting with a
cosmic string of tension $\mu$ is given by \cite{Damour-Vilenkin-moduli}
\beq
S = -\int d^{4} x \left[\frac{1}{2} (\nabla\phi)^{2} + \frac{1}{2}
  m^{2} \phi^{2} + \frac{\sqrt{4\pi}\alpha}{m_{p}} \phi T^{\nu}_{\nu}
  \right] - \mu \int d^{2} \sigma \sqrt{-\gamma}, 
\label{effaction}
\eeq
where $\gamma$ is the determinant of the induced worldsheet metric
$\gamma_{ab} = g_{\mu\nu} X^{\mu}_{,a} X^{\nu}_{,b}$, $X^{\mu}
(\sigma, \tau)$ is the string worldsheet, $T^{\nu}_{\nu}$ is the trace
of the energy momentum tensor of the string, $\alpha$ is the modulus
coupling constant, $m$ is the modulus mass and $m_{p}$ is the Planck
mass.  For $\alpha\sim 1$, the modulus coupling to matter is
suppressed by the Planck scale.  Here, we treat $\alpha$ as a free
parameter and are mainly interested in $\alpha\gg 1$.  Then, the mass
scale characterizing the modulus interactions is $\sim m_p/\alpha \ll
m_p$.  Values as large as $\alpha\sim 10^{15}$ have been discussed in
the literature \cite{Goldberger-Wise}. 

The modulus field equation has the form
\beq
(\nabla^2 - m^2) \phi(x) = - \frac{\sqrt{4\pi}\alpha}{m_{p}} T^\nu_\nu(x),
\eeq
with
\beq
T^\nu_\nu(x) = - 2 \mu \int d\tau d\sigma \sqrt{-\gamma}
\delta^{4}(x^{\alpha} - x^{\alpha} (\sigma, \tau)). 
\label{Tnunu}
\eeq

The power spectrum of modulus radiation from an oscillating loop of
string can be decomposed in Fourier modes as
\cite{Damour-Vilenkin-moduli} 
\beq\label{power-spectrum}
\frac{d P_{n}}{d \Omega} = \frac{G \alpha^2}{2 \pi} \omega_{n} k
|T({\bf k}, \omega_{n})|^2,  
\eeq 
where $G$ is the Newton's constant, $\omega_{n} = \sqrt{k^2 + m^2} = 4 \pi n / L$, $L$ is the length
of the loop,
\beq\label{Tk} 
T({\bf k}, \omega_{n}) = - \frac{4 \mu}{L} \int d^4x \int d\sigma d
\tau \sqrt{-\gamma} \delta^{4}(x^{\alpha} - x^{\alpha} (\sigma, \tau))
e^{ik_\nu X^\nu(\sigma,\tau)} , 
\eeq
and $k^\nu =(\omega_n,{\bf k})$.

We shall be interested in the modulus emission from large loops of
string, having length $L\gg m^{-1}$.  In this case, the characteristic
frequency of loop oscillation is $\omega\sim 1/L \ll m$, so modulus
production is suppressed, except in the vicinity of cusps, where
extremely high frequencies can be reached in a localized portion of
the loop for a brief period of time.  Lorentz factors greater than\footnote{From here on, and until Appendix, we use the notation $\gamma$ only for the Lorentz factor.}
$\gamma$ are reached in a fraction of the loop of invariant length
$\Delta L\sim L/\gamma$. 

The spectrum of resulting particle bursts can be found by expanding
$X^\mu(\sigma,\tau)$ near a cusp
\cite{Peloso-Sorbo,Vachaspati}.\footnote{A detailed analysis of
  particle emission from cusps is given in the Appendix, confirming
  the results obtained in \cite{Peloso-Sorbo,Vachaspati}.} 
One finds that the number of moduli emitted in a single burst with
momenta $k$ in the interval $dk$ (in the center of mass frame of the
loop) is given by 
\beq\label{dNdk}
d {N}(k) \sim \alpha^2 G \mu^2 L^{2/3} k^{-7/3} dk .
\eeq
This distribution applies for $k>k_c$, where
\beq
k_c\sim \frac{1}{4} m\sqrt{mL}.
\label{kc}
\eeq
At smaller $k$ the distribution is strongly suppressed, $dN\approx 0$.

The dominant contribution to the modulus emission comes from the lower
momentum cutoff $k_{min} \sim k_{c}$, so the total number of moduli
per burst is 
\beq
\label{moduli-per-burst}
N \sim \frac{\alpha^2 G \mu^2}{m^2}. 
\eeq

The particles come from a portion of the loop that reaches Lorentz
factors in excess of  
\beq
\gamma_c\sim k_{c}/ m \sim  \frac{1}{4} \sqrt{mL},
\label{gammac}
\eeq
and are emitted into a narrow opening angle $\theta_c$ around the
direction of the string velocity ${\bf v}$ at the cusp,  
\beq
\theta_c\sim \gamma_c^{-1}\sim 4 (mL)^{-1/2}.
\eeq
The total power of modulus radiation can be similarly calculated
as
\beq
P_{m}\sim \alpha^2 G \mu^2 L^{-1/3} k_{c}^{-1/3} \sim \frac{\alpha^2 G
  \mu^2}{\sqrt{mL}}. 
\eeq

The loops also radiate gravitational waves with the power 
\beq
P_{g} \sim \Gamma G\mu^2,
\eeq
where $\Gamma \approx 50$ \cite{Vilenkin-book}. $P_{g} \sim P_{m}$ when $L \sim L_{*}$ which is given by
\beq\label{L_*}
L_{*} \sim \Gamma^{-2} \alpha^{4} m^{-1}.
\eeq
The lifetime of a loop which mainly radiates gravitationally is 
\beq
\tau_{g} \sim \frac{\mu L}{ P_{g}} \sim \frac{L}{\Gamma G \mu},
\eeq
which implies that the characteristic size of the smallest (and most numerous)  loops surviving at time $t$ is
\beq\label{L^g}
L_{min}^{g} \sim \Gamma G \mu t.
\eeq

On the other hand, modulus radiation dominates when $P_{g} \lesssim
P_{m}$ and the loop lifetime is given by 
\beq
\tau_{m} \sim \frac{\mu L}{P_{m}} \sim \frac{L^{3/2} m^{1/2}}{\alpha^{2} G \mu}.
\eeq
The corresponding minimum loop size at time $t$ is
\beq\label{L^m}
L_{min}^{m} \sim \alpha^{4/3} (G\mu)^{2/3} m^{-1/3} t^{2/3}.
\eeq
The transition between the two regimes occurs at
\beq
t_* \sim \frac{\alpha^4}{\Gamma^3 G\mu m}.
\label{t*}
\eeq
Therefore, the minimum loop length is given by (\ref{L^g}) for
$t\gtrsim t_*$ and by (\ref{L^m}) for $t\lesssim t_*$. The
redshift corresponding to $t_*$ is given by
\beq
z_{*} \sim \Gamma^{2} \alpha^{-8/3} (G\mu)^{2/3} (m t_{0})^{2/3}, 
\label{z*-gen}
\eeq
numerically 
\beq
z_* \sim 400  m_{5}^{2/3} \alpha_7^{-8/3} \mu_{-20}^{2/3},
\label{z*}
\eeq
where 
\beq
m_{5}=m/10^{5}GeV,~~~\alpha_7 = \alpha/10^7,~~~ \mu_{-20}=G\mu/10^{-20},
\eeq
and the fiducial values have been chosen anticipating the results in Sec.~IV.




\section{Neutrino production and propagation}

In this section we address some problems in neutrino physics relevant 
for future consideration, namely, the neutrino horizon, the
energy spectrum of neutrinos produced by a modulus decay, the boost of this
spectrum by the cusp Lorentz factor, and some others. 

We start with a note about accuracy of our calculations.  

The main purpose of our work is a discussion of the principle features 
of the phenomenon, i.e., EHE neutrino production by moduli 
from cosmic strings, not an accurate numerical evaluation
of the neutrino fluxes and their detection rates. In particular, our aim 
is to express the results in the form of analytical formulae, so that 
the dependence on input parameters can be easily seen. For this purpose we 
make the following simplifying assumptions. 

We use the CDM cosmological model  with $\Lambda = 0$ and $\Omega_{m} +
  \Omega_{r} = 1$ and use $H_{0} = 72\,km/s Mpc$, $t_{0} = 4.3 \times
  10^{17}\,s$, $t_{eq} = 2.4 \times 10^{12}\,s$, $1+z_{eq} = 3200$,
  the scale factor in the radiation and matter dominated eras are
  $a_{r}(t) \propto t^{1/2}$ and $a_{m}(t) \propto t^{2/3}$. The
  corresponding time-redshift relations are respectively given by
  $(t/t_{0}) = (1+z_{eq})^{1/2}(1+z)^{-2}$ and $(t/t_{0}) =
  (1+z)^{-3/2}.$

For convenience of calculations we assume that at the decay of a 
modulus at rest the neutrino spectrum is $\propto E^{-2}$, while 
in reality this spectrum is not power-law and is approximately 
proportional to $E^{-1.9}$ only for a very large mass of the decaying 
particle.  
 

\subsection{Neutrino horizon}

As they propagate through the universe, the UHE neutrinos $\nu_i$ (or antineutrino 
$\bar{\nu}_i$) with $i=e, \mu, \tau$  are absorbed or loose 
energy in the following three reactions:
\beq
(i)\;\; \bar{\nu}_i + \nu_i \to q_\alpha + \bar{q}_\alpha, ~
(ii)\;\; \nu_i + \bar{\nu}_i \to l+\bar{l},~ 
(iii)\;\; \nu_i + \bar{\nu}_j \to \nu_i + \bar{\nu}_j,  
\eeq
where $q=u_\alpha,~ d_\alpha,~ s_\alpha,~ c_\alpha,~ b_\alpha$ are
quarks with $\alpha=1,~ 2,~ 3$ being color indices and $l= e,~ \mu,~\tau$ 
are lepton flavors. Reactions $(i)$ include only s-channel, and $(ii)$ may 
include both $s$ and $t$ channels. For a rough estimate we can use 
the following approximation for the cross-section,
\beq 
\sigma(s) \approx\left\{ \begin{array}{ll}
(N/\pi)G_F^2s                       ~&{\rm at}~~ s < m_W^2\\*[3mm] 
(N/\pi)G_F^2m_W^2                   ~&{\rm at}~~ s > m_W^2
\end{array}
\right. \;,
\eeq 
where $G_{F} = 1.17\times 10^{-5}$ GeV$^{-2}$ is the
Fermi constant, $s(z) = 2 E_{\nu} m_{\nu} (1+z)$ is the center of mass energy squared
at redshift $z$, $E_\nu$ is the neutrino energy at the present 
epoch, $m_{\nu} \sim 0.1-0.2$ eV is the assumed  neutrino mass, and  
$N \sim 10 - 15$.

UHE neutrinos are absorbed or loose energy in collisions with relic 
background neutrinos whose space number density is 
$n_{\nu} = 56 (1+z)^3$ cm$^{-3}$ and kinetic energy is
$\epsilon_{\nu} = 3.15 T (1+z) = 5.29\times 10^{-4} (1+z)$ eV. 

{\em Neutrino horizon }, i.e., the maximum redshift $z_\nu$, from
which neutrino with the observed energy $E_{\nu}$ can arrive, 
is calculated as 
\beq
\int_{z=0}^{z_{\nu}} dt\; \sigma_{\nu}(z)\; n_{\nu}(z) = 1,
\label{nu_abs}
\eeq
where 
\beq
dt  = \frac{3}{2} t_{0} (1+z)^{-5/2} dz.
\label{dt}
\eeq
At the highest neutrino energies, when $\sigma \sim \sigma_{\max} 
\sim (N/\pi)G_F^2 m_W^2$ , $z_\nu \sim 1.5 \times 10^2$. At energies
below $2\times 10^{11}$~GeV, 
\beq 
z_\nu \sim 2.5 \times 10^2 (E/10^{11}~ {\rm GeV})^{-2/5}.
\label{z_nu}
\eeq
For energies of interest in this paper we shall use $z_\nu \sim 200$
at $E_\nu \gtrsim 10^{20}$~eV. Detectable UHE and EHE neutrinos are produced 
in the matter-dominated epoch.

It is interesting to note that the modulus-string model allows  
to probe the earliest universe with the help of EHE neutrinos: e.g., 
for superconducting strings \cite{BOSV} the maximum redshift is 
$z_{\max} \sim 3$.



\subsection{Modulus decay and  neutrino spectrum}
\label{ModulusDecay}

The rate of decay of a modulus  into the Standard Model (SM) gauge
bosons can be estimated as 
\beq\label{decay-rate}
\Gamma_0 \sim n_{SM} \left(\frac{\alpha}{m_{p}}\right)^{2} m^{3},
\eeq
where $n_{SM} =12$ is the total number of spin degrees of freedom for
all SM gauge bosons, $m$ is the modulus mass, and we assume
interaction of the form \cite{Conlon-Quevedo}
\beq\label{phi-gamma}
\mathcal{L}_{int} \sim \frac{\alpha}{m_p} \phi F_{\mu\nu} F^{\mu\nu} .
\eeq 
The mean lifetime of the modulus in its rest frame is then 
\beq\label{tau_0}
{\tau}_0 \sim 8.1 \times 10^{-17} m_{5}^{-3} \alpha_7^{-2}\,s.
\eeq
For a wide range of parameters $m_5$ and $\alpha$, the lifetimes of
moduli are short even after a large Lorentz boost. In our main scenario, 
the neutrino-producing moduli are born within the neutrino horizon and
decay almost momentarily there. 

However, in principle the redshifts $z > z_\nu$ can also 
contribute to the neutrino flux at $z=0$, if the lifetime of boosted 
moduli is long and they can decay at $z < z_\nu$. In the analysis 
below we argue that such a scenario is disfavored.  

The lifetimes of moduli emitted from cusps are boosted by large Lorentz
factors.  A modulus emitted with a Lorentz factor $\gamma_0=k/m$ at
redshift $z$ and decaying at redshift ${z}_d$ has a lifetime 
\beq\label{tau-z}
\tau(z) \sim \tau_{0} \gamma_0 \frac{1+z_{d}}{1+z}.
\eeq

In order for neutrinos to reach the Earth, they should be produced
within the neutrino horizon at redshifts $z_d\lesssim z_{\nu}$.
Moduli emitted from cusps at $z>z_\nu$ can therefore yield observable
events only if they have large enough lifetime, allowing them to
survive until they reach $z_\nu$.  This gives the condition 
\beq
\tau(z) \sim \tau_{0} \gamma_0 \frac{1+z_{\nu}}{1+z} \gtrsim
t(z_{\nu}) \approx t_{0} (1+z_{\nu})^{-3/2} ,
\eeq
where in the last step we used the fact that in the energy range of
interest $z_{\nu} \lesssim z_{eq}$.  

For $z > z_\nu$  and using $m_5 \gsim 1$ and $\alpha_7 \gsim 1$ 
(which is necessary for detectable neutrino flux; see Sec.~IV), we obtain 
\beq 
\gamma_0 \gtrsim \frac{t_0/\tau_0}{(1+z_d)(1+z_\nu)^{1/2}} \approx 
\frac{3.8 \times 10^{32}}{1+z_d},
\label{gamma0}
\eeq
which is too large at any $z_d$.

In what follows we shall consider only cusp events occurring at $z <
z_\nu < z_{eq}$.  There are no restrictions on the modulus
lifetime in this case, except that it should be short enough for
sufficient fraction of moduli to decay before they reach the Earth.
This is always satisfied in the parameter range of interest.

The decay channel relevant for the neutrino production is the decay
into gluons, via the modulus-gluon interaction of the form
(\ref{phi-gamma}).  The primary gluons from the modulus decay 
initiate the quark-gluon cascade, which turns into hadrons, mostly 
in pions, and then to neutrinos.    

To simplify calculations and analysis, we shall assume the neutrino 
production spectrum $dN/dE \propto E^{-2}$, close to the power-law 
approximation $E^{-1.9}$ obtained for a large mass of the decaying particle  
using Monte Carlo simulation and DGLAP method \cite{DGLAP}.

The neutrino spectrum from a modulus  at rest is then
\beq\label{fragmentation}
\frac{dN^*_\nu}{dE_*} \equiv \xi^*_{\nu} (E_*,m) \approx 
\frac{1}{2} f_\pi b_* \frac{m}{E_*^{2}},
\eeq
where $m$ is modulus mass, $E_*$ is neutrino energy,
$b_*$ is given in terms of the ratio of maximum $\eps_*^{\max} \sim 0.1 m$  
and minimum 
$\eps_*^{\min}$ neutrino energy,
\beq
b_*=[\ln(\eps_*^{\max}/\eps_*^{\min})]^{-1} ,
\label{ln*}
\eeq
and $f_\pi \sim 1$ and 1/2 are the fractions of energy transferred from
the modulus to pions and from pions to neutrinos, respectively. 

The spectrum of low-energy 
neutrinos is a model-dependent feature, but generically suppression of
this spectrum is provided by suppression of soft gluon emission   
due to coherence effect in the parton cascade. It results in the
Gaussian peak in the spectrum of pions, parents of neutrinos. We describe 
effects of low-energy suppression of neutrino spectrum introducing 
formally the energy $\eps_*^{\min}$ in Eq.(\ref{ln*}), where the suppression starts, and refer to it as ``the minimal energy".   Determining the value of $\eps_*^{min}$ would require numerical calculations.  Here, we shall parametrize 
\beq
\varepsilon=\epsilon_*^{\min}/1~GeV.   
\label{param}
\eeq

\subsection{Lorentz boost and beaming}
\label{boost}

Emerging from a cusp segment, a modulus obtains very large Lorentz 
factor $\gamma$ corresponding to the point of exit. The typical 
Lorentz factor, as it is calculated below, reaches 
$\gamma \sim 10^{12} - 10^{13}$. The neutrino energy $E_*$ 
in the modulus rest system is boosted as
\beq
E = \gamma E_*(1+\beta \cos \theta_*),
\label{L-transform}
\eeq
where $\beta=v/c$, and $\theta_*$ is the angle between the directions of neutrino motion    
and of the boost in the rest frame of the modulus. 

First of all we calculate how the neutrino spectrum (\ref{fragmentation})
changes under the transformation (\ref{L-transform}). For this we use the 
conservation of the number of particles before and after the Lorentz boost: 
\beq
dN_*(E_*,\theta_*)=\frac{1}{2} b_* m \frac{dE_*}{E_*^2} 
\frac{d\Omega_*}{4\pi} = dN(E,E_*).
\label{conserv}
\eeq
Using 
\beq
d\Omega_*=2\pi\; d\cos \theta_* = 2\pi \frac{dE}{\gamma \beta E_*}, 
\label{dOmega}
\eeq
we obtain in terms of the new variables $E$ and $E_*$ : 
\beq
\frac{d^2N}{dE dE_*}= b_* \frac{m}{4\gamma \beta E_*^3}.
\label{double-spec}
\eeq
Integrating with respect to $dE_*$ we have 
\beq 
\frac{dN}{dE} = b_* \frac{m}{8\gamma \beta}\frac{1}{E_{*\min}^2(E)},
\label{spectr1}
\eeq
where the minimum neutrino energy for a fixed $E$ is  
\beq 
E_{*\min}(E) = \left\{ \begin{array}{ll}
E/[\gamma (1+\beta)]  ~&{\rm if}~~ E \geq \gamma (1+\beta)\eps_*^{\min} \\*[3mm] 
\eps_*^{\min}         ~&{\rm if}~~ E \leq \gamma (1+\beta)\eps_*^{\min}
\end{array}
\right. \;,
\label{Emin*}
\eeq 
Finally, we obtain for $\beta \approx 1$ and for decay at epoch $z$  
\beq
\frac{dN}{dE} \equiv \xi_\nu (E,k,z)= \frac{1}{2}\frac{b_*}{1+z}\frac{k}{E^2},
\label{spectrum}
\eeq
if the neutrino energy at the present epoch is  
$E \geq \gamma (1+\beta)\eps_*^{\min} $.
For $E \leq \gamma (1+\beta)\eps_*^{\min} $,~~
$dN/dE$ does not depend on the neutrino energy $E$ : 
\beq
\frac{dN}{dE} \equiv \xi_\nu(E,k,z)=\frac{1}{8}\;\frac{b_*}{1+z}\; 
\frac{k}{\gamma^2}\; \frac{1}{(\eps_*^{\min})^2}.  
\label{spectrum1}
\eeq
In both formulae above, $k=\gamma m$ is the modulus energy. 

We now briefly discuss the effect of beaming. Strong beaming of the produced particles is a remarkable feature of the
cusp models, which provides interesting observational consequences. 
For moduli with the range of masses considered here the beaming is very strong (see Section \ref{ModulusRadiation}).  When gravitational 
radiation dominates, the Lorentz factor at $z=0$ can be estimated as 
$\gamma_c \approx 4.5 \times 10^{13}~m_5^{1/2}\mu_{-20}^{1/2}$.
When modulus radiation dominates, the Lorentz factor at $z_*$ is 
$\gamma_c=\alpha^2/4\Gamma=5\times 10^{11} \alpha_7^2$. As a typical Lorentz 
factor for neutrino production one may consider that at neutrino
horizon $z_\nu \sim 200$,~~ 
$\gamma_c = 8.5\times 10^{11} m_5^{1/2}\mu_{-20}^{1/2}$.

In the frame where the modulus is at rest, neutrinos are emitted isotropically.
After a Lorentz boost, most of them move within a narrow cone with 
$\theta \sim 1/\gamma$ in the direction of the boost, with energies 
$E \sim \gamma E_*$. However,
neutrinos which are emitted in the rest frame within a narrow cone with
$\theta_{*c}'=1/\gamma$ in the backward direction are moving after the
boost within a wide angle $\theta \sim \pi/2$ in the  direction opposite 
to the narrow high-energy jet and typically have very low energies 
$E \sim E_*/\gamma$. The total number of these neutrinos is $4\gamma^2$
times smaller than that in the high-energy jet, and they are undetectable because
of their small number and low energies.  The typical neutrino energies in 
high-energy beam is $E \sim \gamma E_*$, but low-energy neutrinos with 
$E \sim E_*/\gamma$ are also present there. 

Formally, the minimum neutrino energy is extremely low, 
$E \sim \eps_*^{min}/\gamma$, but the number of such neutrinos is very small. 
However, when neutrino spectrum is changing from $E^{-2}$ at high 
energies to $dN/dE \propto const$, as in Eq.~({\ref{spectrum1}), the 
detectability of neutrinos is sharply decreased, and thus 
$E_{\min} \approx \gamma (z) \epsilon_*^{\min}$  
can be regarded as an effective low-energy end of the spectrum at 
epoch $z$.  
To estimate the low-energy spectrum cutoff for neutrinos generated at 
epoch $z_\nu \sim 200$ and observed now, we use the parametrization 
(\ref{param})  
and our estimate for the Lorentz factor at $z_\nu \sim 200$, $\gamma_c=8.5\times 10^{11} m_5^{1/2} \mu_{-20}^{1/2}$.  Taking into account the redshift 
of the neutrino energy, we find
$E_{\min} \sim 4.3\times 10^{9} \varepsilon m_5^{1/2}\mu_{-20}^{1/2}$~GeV.
This estimate demonstrates the remarkable feature of our model:
the predominant generation of the highest energy neutrinos.  A more
realistic estimate of $E_{\min}$ for the diffuse neutrino flux will 
be made in Section \ref{NeutrinoFlux}. 

EHE neutrinos propagate as a jet in a cone with an opening angle 
$\theta \sim 1/\gamma$. The duration of the neutrino pulse is very short, 
$\tau \sim L/\gamma^3$, and all neutrinos reach the detector almost 
simultaneously, due to the smallness of the neutrino mass. The effective area  
illuminated by arriving neutrinos is much larger than the area controlled 
by the detector, but simultaneous appearance of two-three showers 
in the field of view of a large detector is possible for some
parameter values (see \cite{BOSV} for discussion and calculations). 
    
Beaming is a property of all particles emitted from 
a cusp, in particular gamma-rays. In some models part of observed GRBs 
may be produced by cosmic strings \cite{Berezinsky-Hnatyk-Vilenkin}. 
Since the total number of strings in the Milky Way is tremendously large 
($N \sim 10^9$), one may expect that UHE GRBs from the Milky Way may be 
observable. In fact, the rate of predicted bursts is too strongly
suppressed by the beaming factor $\Omega/4\pi = 1/(4\gamma^2)$
to be detectable. The backward component is distributed within a wide 
solid angle and is not suppressed by the this factor, but the total 
number of photons and their energies are too small for detection.

    
\section{Neutrino Bursts from Moduli}


\subsection{Loop Distribution}

The predicted flux of EHE neutrinos depends on the typical length of
loops produced by the string network.  The typical loop sizes have been a subject of much
recent debate, with different simulations \cite{BB,AS,Hindmarsh,MartinsShellard,Ringeval,Vanchurin-Olum-Vilenkin,Olum-Vanchurin,Jose-Ken-Ben} and analytic
studies \cite{Polchinski-Rocha,Dubath} yielding different answers.  Here we shall adopt the
picture suggested by the largest, and in our view, the most accurate
string simulations performed to date
\cite{Jose-Ken-Ben}. According to this
picture, the characteristic length of loops formed at cosmic time $t$
is given by the scaling relation 
\beq
L \sim \beta t,
\eeq 
with $\beta \sim 0.1$.   

The number density of loops with lengths in the interval from $L$ to
$L+dL$ can be expressed as $n(L,t)dL$.  Of greatest interest to us are
the loops that formed during the radiation era $(t<t_{eq})$ and still
survive at $t>t_{eq}$.  The density of such loops is given by
\beq
\label{matt-loop-density}
n(L,t) dL \sim p^{-1} \zeta (\beta t_{eq})^{1/2} t^{-2} L^{-5/2} dL ,
\eeq
where $p$ is the string reconnection probability and $\zeta\sim 16$ is
the parameter characterizing the density of infinite strings with
$p=1$, $\rho_\infty = \zeta\mu/t^2$.   

The dependence of the loop density on $p$ is somewhat uncertain and
can only be determined by large-scale numerical simulations.  Here we
have adopted the $p^{-1}$ dependence suggested by simple arguments in, e.g.,
\cite{Damour05,Sakellariadou}.   
The reconnection probability is $p=1$ for ordinary cosmic strings.
Its value for F- and D-strings of superstring theory has been
estimated as \cite{Jackson-Jones-Polchinski}  
\beq
10^{-3}\lesssim p \lesssim 1.
\eeq


The distribution (\ref{matt-loop-density})
applies for $L$ in the range from the minimum length
$L_{min}$ to $L_{max}\sim \beta t_{eq}$.  The lower cutoff $L_{min}$
depends on whether the energy dissipation of loops is dominated by
gravitational or by modulus radiation. It  
is given by (\ref{L^g}) for $z<z_*$ and by (\ref{L^m}) for $z_*
  <z<z_{eq}$, with $z_*$ from Eq.(\ref{z*}).  For $z_*>z_{eq}$,
  the dominant energy loss is gravitational radiation and
  Eq.(\ref{L^g}) for $L_{min}$ applies in the entire range of
  interest. 

The string motion is overdamped at early cosmic times, as a result of
friction due to particle scattering on moving strings.  The overdamped
epoch ends at \cite{Vilenkin-book} 
\beq
t_{damp} \sim (G\mu)^{-2}t_p ,
\eeq
where $t_p$ is the Planck time.  In the above analysis we have assumed
that loops of interest to us are formed at $t>t_{damp}$.  The corresponding
condition is 
\beq
L_{min}(t)\gtrsim \beta t_{damp} .
\label{friction}
\eeq
This bound assumes that
the strings have non-negligible interactions with the standard model
particles, so it may not apply to F- or D-strings of superstring theory.  In any case, we have verified that (\ref{friction}) is
satisfied for parameter values that give a detectable flux of
neutrinos. 
\subsection{Gravity- and moduli-dominated regimes. Restrictions
imposed by $z_*$}
\label{restrictions}

The value of $z_*$ given by Eq.~(\ref{z*}) marks the boundary between 
two regimes of string evolution.  In the first regime, at $z<z_*$, the 
string energy loss is dominated by gravitational radiation.  This
includes 
the entire relevant range of $z$ for $z_* >z_\nu$ and the range    
$0 \leq z \leq z_*$ for $z_*<z_\nu$.  We shall call it the 
{\em gravity-dominated regime}.  

The second regime is dominated by the modulus radiation; we shall call 
it the {\em moduli-dominated regime}.  It corresponds to the redshift 
interval $z_* \leq z \leq z_\nu$ and exists only when $z_* < z_\nu$.

It is often convenient to perform the calculations for the fixed  
value of $z_*$, considering it as a free parameter. In this case the 
space of three physical parameters $\alpha_7, m_5$ and $\mu$ 
is restricted by Eq.~(\ref{z*}) as 
$400 \alpha_7^{-8/3} m_5^{2/3} \mu_{-20}^{2/3}=z_*$, which we will use
in the form 
\beq
\mu_{-20}=(z_*/400)^{3/2}\alpha_7^4\;m_5^{-1}. 
\label{muz*}
\eeq
Then our calculated quantities, such as neutrino fluxes and characteristic
energies, will depend on two parameters, 
$\alpha_7$ and $m_5$, and the fixed value of $z_*$. 

The value of fixed $z_*$ characterizes ``the model". One should  
distinguish two major classes of models: with high $z_* > z_\nu$ , 
which corresponds to gravity-dominated regime, and with low 
$z_* < z_\nu$, which includes both the gravity-dominated regime 
at $0 \leq z \leq z_*$, and the moduli-dominated regime at 
$z_* \leq z \leq z_\nu$.


\subsection{Neutrino Flux}
\label{NeutrinoFlux}

As we argued in Section \ref{ModulusDecay}, the neutrino-producing  
moduli are born and decay 
within the neutrino horizon at $z \leq z_\nu$. We shall first 
consider the high $z_*$ models with $z_* > z_\nu$ (gravity-dominated regime)
and then study the case $z_* < z_\nu$, which includes both  
the moduli-dominated regime at $z_* \leq z \leq z_\nu$ and the
gravity-dominated regime at $0 \leq z \leq z_*$. 


The neutrino flux can be most generally calculated as  
\beq
J_{\nu}(E,z) = \frac{1}{4\pi} \int \frac{dV(z)}{1+z}\; d\dot{N}_b\; dN_X^b (k)
\;\frac{\Omega_k}{4\pi}\; \frac{1}{\Omega_k r^2(z)}\; \xi_\nu (E,z,k),    
\label{Jnu}
\eeq
where the proper volume for the matter-dominated epoch is  
\beq
dV(z) = 54 \pi t_{0}^3 [(1+z)^{1/2} -1]^{2} (1+z)^{-11/2} dz\; ,
\label{dV(z)}
\eeq
the rate of bursts is $d\dot{N}_b= n(L)dL/(L/2)$ with $n(L)dL$ being 
the number density of loops with length $L$ in the interval $dL$ (see 
Eq.~(\ref{matt-loop-density})), the number
of moduli $dN_X^b$ emitted in a burst with momenta $k$
in the interval $dk$ is given by Eq.(\ref{dNdk}), 
$\Omega_k/4\pi$ is the probability that a randomly oriented 
burst is directed to the observer, $\Omega_k r^{2}(z)$ is the 
area of the irradiated spot at the observer's location, and  
$\xi_\nu (E,z,k)$ is the spectrum of neutrinos from decay of a modulus 
with momentum $k$, given by Eqs.~(\ref{spectrum}) and (\ref{spectrum1}).

Integration in Eq.~(\ref{Jnu}) goes over $k$, $L$ and $z$. For
integration over $k$ and $L$ only lower limits are essential, and
they are given by $k_{\min}=m\sqrt{mL}$ and $L_{\min}$ from 
Eq.~(\ref{L^g}) or Eq.~(\ref{L^m}) for the gravity-dominated and 
moduli-dominated regimes, respectively.
In the case $z_* > z_\nu$, we have $z_{\max} = z_\nu$, while 
$z_{\min}$ is determined by the rate of bursts or by minimum energy of neutrino $E_{\min}(z)$ at epoch $z$ 
as explained below. 

Consider first the limit $z_{\min}$ imposed by the rate of bursts.
The average rate of bursts ${\dot N}_b(<z)$ that occur in the redshift 
interval between 0 and $z$ is a growing function of $z$, and we can 
define $z_b$ as the redshift at which this rate is a few bursts per year.  
No bursts will be detected from $z\ll z_b$, so we should introduce a
lower 
cutoff of $z$-integration at $z_{min}=z_b$.
The rate of burst $\dot{N}_b$ is calculated 
in Section \ref{RateBursts} and $z_b$ is found to be small in the parameter 
range that we are considering here.  Hence, the condition $z>z_b$ does
not yield a significant constraint for the $z$-integration in Eq.~(\ref{Jnu}). 

Another constraint to consider is that the energy $E$ of neutrinos should 
be above the minimal energy, $E>E_{\min}(z)=\gamma_c (z) \eps_*^{min}/(1+z)$, 
where
$\gamma_c(z)= \frac{1}{4}[m L_{\min}(z)]^{1/2}$ is the characteristic 
Lorentz factor for loops of minimal length $L_{\min}(z)$ at epoch $z$.  
For the gravity-dominated regime we are considering here
\beq
\gamma_c(z) = 4.5 \times 10^{13} 
m_5^{1/2}\mu_{-20}^{1/2} (1+z)^{-3/4}.
\label{gamma_c}
\eeq 
To proceed, it will be convenient to use $z_*$ in place of the string 
tension $\mu$ as a free parameter.  With the aid of Eq.~(\ref{muz*}) we have
\beq
E_{\min}(z) =  E_0 \alpha_7^{2} (z_*/z_\nu)^{3/4} (1+z)^{-7/4},
\label{Eminz}
\eeq 
where
\beq
E_0 = 2.7\times 10^{13}\varepsilon ~{\rm GeV},
\eeq
and $\varepsilon$ is the parametrization factor introduced in 
Eq.~(\ref{param}). 
The lower bound of $z$-integration $z_{min}$ can now be found as the 
value of $z$ for which $E_{min}(z)=E$,
\beq
1+z_{min}(E)=(E_0/E)^{4/7} \alpha_7^{8/7}(z_*/z_\nu)^{3/7}.
\eeq
Integrating Eq.~(\ref{Jnu}) over $z$
from $z_{min}(E)$ to $z_\nu$, since we assumed above $z_*>z_\nu$,  we obtain
\beq
E^2 J_\nu(E) = 2.5\times 10^{-9} p^{-1}\alpha_7^2 m_5^{-1/2}
\left (\frac {z_\nu}{200}\right )^{1/2}
\left[ 1-\left(\frac{1 + z_{min}(E)} {1+z_\nu}\right)^{1/2}\right]~ 
{\rm GeV/cm}^2{\rm s~sr}.
\label{E2Jnunew}
\eeq

The calculated flux for ``normalizing" set of parameters $p=1$
(ordinary strings) and $\alpha_7=m_5=z_*/z_\nu=1$ is shown in 
Fig.~\ref{fig1} by curve ``theor.3". This flux is low and detectable 
only by SKA. The largest flux in Fig.~\ref{fig1} is presented by curve
``theor.1" for the parameters $p=1,\; \alpha_7=2,\; m_5= 0.1,\;  
z_*/z_{\nu}=1$ . It is close to upper limit shown by curve ``$E^{-2}$ 
cascade", and is detectable by JEM-EUSO, LOFAR and SKA. Here and
everywhere below  we assume $\varepsilon =1$. 

The maximum energy of neutrinos is 
determined by $E_{\max} \sim 2 \gamma \epsilon_*^{\max}$ at 
generation and can be extremely large, but the flux of these neutrinos 
is suppressed as $E^{-2}$.

We shall now consider the low $z_*$ models with $z_* \leq z_\nu$,
and  calculate first the neutrino flux generated in the redshift interval
$z_* \leq z \leq z_\nu$, where energy losses are 
moduli-dominated.  Then we calculate  flux in the
interval $0 \leq z \leq z_*$, where 
the gravitational radiation dominates. 

For the interval  $z_* \leq z \leq z_\nu$ and fixed $z_*$, 
the neutrino flux is calculated using Eq.~(\ref{Jnu}) and 
the parameter restriction 
in the form $\mu_{-20}^{2/3}=(z_*/400)\alpha_7^{8/3} m_5^{-2/3}$.
As a result we have 
\beq 
E^2 J_\nu (E)= \frac{1}{2} K p^{-1}\; \alpha_7^2\; m_5^{-1/2} (1+z_*)
\int_{z_{\min}(E)}^{z_\nu} dz\; (1+z)^{-3/2},
\label{E^2Jnu2}
\eeq
with $K=1.8\times 10^{-10}$~GeV/cm$^2$s sr.

The lower limit of integration in Eq.~(\ref{E^2Jnu2}) is obtained as 
above from $E_{\min}(z)=\gamma_c(z)\varepsilon_{\min}/(1+z)$. Using the 
condition $E \geq E_{\min}(z)$ and $z_{\min} \geq z_*$ one obtains  
\begin{equation}
z_{\min}(E)=\left\{ \begin{array}{lll}
z_*                       ~&{\rm at}~~ E \geq \tilde{E}_{\min}\\ 
(E/E_0)^{-2/3}\alpha_7^{4/3}z_*^{1/3} ~&{\rm at}~~ E_{\min}\leq E 
\leq \tilde{E}_{\min},\\
z_\nu                     ~&{\rm at}~~ E \leq E_{\min}
\end{array}
\right. 
\label{Emin-mod-regime}
\end{equation}
with $E_{\min}=1.3\times 10^{18}\varepsilon \alpha_7^2 (z_*/50)^{1/2}$~eV,
$\tilde{E}_{\min}=1\times 10^{19}\varepsilon \alpha_7^2 (50/z_*)^2$~eV, 
and $E_0=5\times 10^{20} \varepsilon$~eV. Finally we have 
\beq
E^2 J_\nu (E)= K p^{-1}\; \alpha_7^2\; m_5^{-1/2}\frac{1+z_*}
{(1+z_\nu)^{1/2}}\left[\left(\frac{1+z_\nu}{1+z_{\min}}\right)^{1/2}-1\right].
\label{E2Jnu3}
\eeq
Using in Eq.~(\ref{E2Jnu3})  $z_{\min}(E)=z_*$ at $E \geq \tilde{E}_{\min}$
(see Eq.~(\ref{Emin-mod-regime})) one finds the high energy (HE) asymptotic
of the flux in the moduli-dominated regime
\beq 
E^2 J_\nu (E)= 1.8 \times 10^{-10} p^{-1} \alpha_7^2 m_5^{-1/2} 
(1+z_*)^{1/2}\left [1- \left (\frac{1+z_*}{1+z_\nu}\right )^{1/2}\right ]
~ {\rm GeV~ cm}^{-2}~{\rm s}^{-1}~ {\rm sr}^{-1},
\label{E2JnuHE}
\eeq
valid at $E \gsim \tilde{E}_{\min}$.

We have to add also the neutrino flux generated in
the interval $0 \leq z \leq z_*$ where the 
gravitational energy losses dominate. In this case $L_{\min}$ is given by 
Eq.~(\ref{L^g}). For HE asymptotic one finds 
\beq
E^2 J_\nu (E)= 1.8 \times 10^{-10} p^{-1} \alpha_7^2 m_5^{-1/2} 
(1+z_*)^{1/2}~ {\rm GeV~ cm}^{-2}~{\rm s}^{-1}~ {\rm sr}^{-1}.
\label{E2Jnu4HE}
\eeq
For an easily understandable reason, the
HE asymptotic here is $(z_*/z_\nu)^{1/2}$ lower 
than in the case $z_* > z_\nu$ given by Eq.~(\ref{E2Jnunew}).  
Less trivial is the coincidence of HE
asymptotic in the gravitational regime $(0,\; z_*)$  given by 
Eq.~(\ref{E2Jnu4HE}) and in the moduli-dominated regime $(z_*,\; z_\nu)$ 
given by Eq.~(\ref{E2JnuHE}). It is explained by the fact that 
dominant contributions in both cases are given by epochs with redshift 
$z_*$.  The general conclusion about neutrino production is therefore 
increasing the flux with growth of $z_*$ until it reaches $z_\nu$. 
Thus, the high $z_*$ models predict the largest neutrino fluxes. 

Apart from this, more efficient neutrino detection in high $z_*$ models,
follows from a lower cutoff energy $E_{\min}$. Indeed,
the minimal energy of neutrinos for both 
regimes $(0.\; z_*)$ and $(z_*,\;z_\nu)$ is determined by 
the same expression
\beq
E_{\min}(z_*)=\frac{\gamma_c(z_*)}{1+z_*}\varepsilon_*^{\min}= 
\frac{1}{4}\frac{\alpha^2}{\Gamma}\frac{\varepsilon_*^{\min}}{1+z_*},
\label{Emin-both}
\eeq
while for the gravity-dominated regime with $z_* > z_\nu$ it is 
determined by $z_\nu$ as 
\beq
E_{\min}(z_\nu)=\frac{\gamma_c(z_\nu)}{1+z_\nu}\varepsilon_*^{\min} .
\label{Emin-grav1}
\eeq
Increasing $z_*$ in Eq.~(\ref{Emin-both}) we decrease $E_{\min}$, which 
is favorable for detection by JEM-EUSO.  

One may also see that moduli-dominated regime gives subdominant  
effect as compared with gravity-dominated one at $z_* > z_\nu$. 

\subsection{The Rate of Bursts}
\label{RateBursts}

The rate of bursts is not a physically measured quantity, but it can
serve as indicator of detectability of the burst radiation.

The rate of cusp bursts with their cones of radiation directed to the
observer is given by 
\beq
\dot{N}_b= \int \frac{dV(z)}{1+z}\; \frac{n(L,z)}{L/2}dL\; 
\frac{\Omega}{4\pi},
\label{burst-rate-gen}
\eeq
where $dV(z)$ and $n(L)dL$ are the same as in Eq.~(\ref{Jnu}). 
The quantity $\Omega/4\pi$ gives the probability for the observer 
to be located within the cone of the cusp radiation. Such a location does not 
guarantee detection of this radiation, because the area of 
irradiated spot $\Omega r^2$ is much bigger than the size of the Solar 
system, but too low rate of bursts means that the signal is undetectable.  
To calculate $\Omega \approx \pi \theta^2$ we use 
$\theta \sim (k_cL)^{-1/3}$ (see the discussion above 
Eq.~(\ref{theta_k}) in the Appendix for details). 

We calculate the rate of bursts for  $z < z_*$,
when
gravitational radiation dominates. The rate integrated from 
$z=0$ up to redshift $z$ is given by 
\beq
\dot{N}_b(\leq z) = \frac{54\pi}{7}2^{4/3}\; p^{-1}\zeta\beta^{1/2}\; 
\frac{(t_{eq}/t_0)}{(\Gamma G\mu)^{7/2}}\; \frac{1}{mt_0^2}\;  I(z)=
1.1 \times 10^7\;  \mu_{-20}^{-7/2}\; m_5^{-1}\; I(z)\; {\rm yr}^{-1},
\label{Nb}
\eeq
where 
\beq 
I(z)= \int_0^{z}dz' [(1+z')^{1/2} - 1]^2 (1+z')^{7/4}.
\label{int}
\eeq
Numerical values of $I(z)$ are shown in Table I.

\begin{table}[h!]
\begin{center}
\caption{Numerical values of integral $I(z)$ }
\vspace{5mm}
\label{table1}
\begin{tabular}{|c|c|c|c|c|}
\hline
$z$  & $0.1$ & $0.5$ & $1.0$ & $2.0$ \\
\hline
$I(z)$ & $9.12\times 10^{-5}$ & $0.015$ & $0.165$ & $1.99$ \\
\hline
\end{tabular}
\end{center}
\end{table}

In Section \ref{NeutrinoFlux} $z_{\min}\equiv z_b$ is defined from 
condition $\dot{N}_b(\leq z_b)$ is a few per year. 
From Table \ref{table1} and Eq.~(\ref{Nb}) one can see that the rate of 
bursts is large enough even at redshift as small as 0.1, 
and thus the condition $z > z_b$ does not impose a significant 
constraint for the integration of Eq.~(\ref{Jnu})  over $z$.

\subsection{Cascade Upper Bound}
\label{CascadeBound}

An upper bound on the neutrino flux follows from the observed diffuse 
HE gamma-ray
background, since neutrino production via pion/kaon decays is
accompanied by high energy electron and photon production. These
electrons and photons, interacting with CMB and Extragalactic
Background Light (EBL) photons, produce an electromagnetic cascade,
whose energy density $\omega_{cas}$ must not exceed that  
of the observed diffuse gamma radiation. This results in the upper 
limit on the diffuse neutrino flux \cite{cascade-upper-limit}. 
With the assumption of $E^{-2}$ generation spectrum of neutrinos
the cascade upper limit can be written as
\beq 
E^2 J_\nu (E)\leq \frac{c}{4\pi} \frac{\omega_{\rm cas}^{\max}}
{\ln (E_{\max}/E_{\min})},
\label{cascade-up-lim}
\eeq
where $E_{\max}$ and $E_{\min}$ are the maximum and minimum neutrino energies,
respectively, and $\omega_{\rm cas}^{\max}$ is the maximum cascade energy 
density allowed by observation of the isotropic diffuse gamma-radiation.  

According to recent Fermi-LAT observations \cite{Fermi} this energy 
density is $\omega_{cas}^{\max} = 5.8 \times 10^{-7}\, {\rm eV/cm^{3}}$, 
as it follows from the analysis \cite{BGKO,Ahlers}.
This limit results from a comparison of the cascade 
spectrum at 100 GeV with the Fermi data, while the lower energies give 
a weaker limit. For our case we assume that cosmological epochs with 
redshifts $z \geq z_{cas}$, when gamma-rays with $E_\gamma > 30$~GeV 
are absorbed, 
make a negligible contribution to the obtained upper limit. One can
estimate $z_{cas}$ in the following way.  At $z=0$ a very sharp absorption
of gamma-rays on CMB occurs at $E_\gamma^{abs}(0) \approx 100$~TeV. 
At epoch $z$ this energy is $(1+z)$ lower. Taking into account the redshift 
of these photons, we estimate $z_{cas}$ as 
$z_{cas} \sim (E_\gamma^{abs}/E_\gamma)^{1/2} \sim 60$.  

The energy density for the electromagnetic cascade radiation resulting 
from modulus decays can be expressed as
\beq
\omega_{cas} = \frac{1}{2} f_\pi \int \frac{dt}{(1+z)^{4}} 
\frac{ n(L, z) dL}{L/2} N(k) k dk,
\label{omega-cas-gen}
\eeq
where $f_\pi \sim 1$ and 1/2 are the fractions of energy transferred from the modulus 
to pions and from pions to electrons and photons, respectively, $dt$ is
given by Eq.~(\ref{dt}), the density of the loops
$n(L)$ is given by  Eq.~(\ref{matt-loop-density}), and the number of 
moduli emitted per burst $N(k)$ is given by Eq.~(\ref{dNdk}). The limits of 
integration in Eq.~(\ref{omega-cas-gen}) are as follows : 
$z_{\min}=0$, $z_{\max}=z_{cas} \sim 60 $, $k_{\min}=(1/4)m\sqrt{m L}$,
and $L_{\min}$ is given by Eqs.~(\ref{L^g}) and (\ref{L^m}) for the 
cases $z_* > z_{cas}$ and  $z_* < z_{cas}$, respectively. 

Consider first the case $z_* > z_{cas}$, when gravitational radiation
dominates. Performing the integration we obtain 
\begin{eqnarray}
\omega_{cas} &=& 9\times 2^{-1/3} p^{-1} \zeta \beta^{1/2} \Gamma^{-2} 
\frac{(t_{eq}/t_0)^{1/2}}{(t_0 m)^{1/2}} \alpha^2 
\frac{m_{Pl}^2}{t_0^2} (1+z_{cas})^{1/2}, \nonumber\\
&=& 5.8\times 10^{-9} p^{-1} \alpha_7^2 m_5^{-1/2} (z_{cas}/60)^{1/2}\; 
{\rm eV/cm}^3.
\label{cas-density1}
\end{eqnarray}

For  $ z_* < z_{cas}$ we have to integrate over the interval 
$0 \leq z \leq z_*$, where gravitational radiation dominates, and over  
interval $z_* \leq z \leq z_{cas}$, where moduli radiation prevails. 
For the first case (gravity-dominated regime at $0 \leq z \leq z_*$ ) one
obtains 
\beq
\omega_{cas}= 4.7\times 10^{-9} p^{-1} \alpha_7^2\; m_5^{-1/2}(z_*/40)^{1/2}~ 
{\rm eV/cm}^3 .
\label{cas-density2}
\eeq
For the second case (moduli-dominated regime at $z_* \leq z \leq z_{cas}$)
we have using Eq.~(\ref{muz*}) 
\beq
\omega_{cas}= 4.7\times 10^{-9} p^{-1} \alpha_7^2\; m_5^{-1/2}(z_*/40)^{1/2}
[1- (z_*/z_{cas})^{1/2}]~ {\rm eV/cm}^3 .
\label{cas-density3}
\eeq
Note that this is the same as (\ref{cas-density2}), apart from the 
last factor.  The reason is that in both cases the main contribution 
to $z$-integration comes from $z\sim z_*$,  like in case already 
discussed for neutrino fluxes.

For reasonable values of $z_*$, all three $\omega_{cas}$ given by 
(\ref{cas-density1}) - (\ref{cas-density3}) are less than maximally 
allowed $\omega_{cas}^{\max} = 5.8 \times 10^{-7}\,$~eV/cm$^{3}$. 


Two remarks about the cascade limit for our model are now in order.  

This limit is not strictly enforced because cascade production occurs at very
large redshifts, while
$\omega_{cas}$ is constrained mainly due to the highest 
energy cascade photons with $E \approx 100$~GeV (see Fig.~1 in 
\cite{BGKO}). These photons are
absorbed by EBL at earlier cosmological epochs, and therefore in cosmic
string models, where the main part of the cascade energy density is
produced at large redshifts, the constraint on $\omega_{cas}$ is weaker and 
values higher than  $5.8 \times 10^{-7}$~eV/cm$^{3}$ are allowed. 
Therefore, neutrino fluxes above $E^{-2}$-cascade upper limit
taken from \cite{BGKO} and shown in Fig.~\ref{fig1} are not necessarily excluded.  

One may also expect that beaming of cascades may reduce the efficiency 
of the cascade limit. The calculation of beam widening in the
ambient magnetic field shows that narrow beaming survives only for very 
weak magnetic fields of order $10^{-15}$~G.


\subsection{Superstring-motivated example for high $z_*$ model with 
$z_* > z_\nu$}
\label{example1}

Because of the simplifying assumptions adopted in this paper, 
we cannot reliably determine the domain in the parameter space of  
$\alpha$, $m$ and $G\mu$, which yields detectable fluxes of EHE
neutrinos.

Instead, we shall discuss some illustrative examples of parameter 
choices, for which neutrino detection in future experiments is possible. 
Among such projects we shall consider detectors aimed at the 
highest energy neutrinos above $10^{20}$~eV: JEM-EUSO \cite{JEM}, 
Anita II \cite{anita}, LOFAR \cite{Haungs,lofar},
SKA\cite{Haungs,ska}, and LORD \cite{lord}. JEM-EUSO is based on space
detection of fluorescent light from the showers in the  atmosphere. 
All others are based on radio detection of the showers due to the Askarian 
effect \cite{Askarian}.  The sensitivities of three of these experiments,
JEM-EUSO (expected to be launched in 2015), LOFAR and SKA, are shown  
in Fig.~\ref{fig1}.

First we shall consider a superstring-motivated example for cosmic strings 
evolving in the gravity-dominated regime, assuming that $z_* > z_\nu$. 
We fix plausible parameters, both for strings and for particle physics.   
Specifically, we consider the large volume string
compactification model \cite{large-volume2}, which is characterized by
an intermediate string scale $m_s\sim 10^{11}~GeV$ and a TeV-scale
supersymmetry (SUSY) breaking.  The hierarchy between the Planck and
SUSY breaking scales in this model is due to an exponentially
large volume of the compact extra dimensions, $V_{comp}\equiv
{\mathcal V}l_s^6$, where ${\mathcal V}\sim 10^{15}$ and $l_s\sim
m_s^{-1}$ is the string length scale.

Apart from the volume modulus, which has gravitational strength 
couplings to ordinary matter, the other Kahler moduli have large 
couplings of the order \cite{Conlon-Quevedo}
\beq
\alpha \sim \sqrt{\mathcal{V}}.
\eeq
With $\mathcal{V} \sim 10^{15}$, we have $\alpha\sim 10^{7.5}$.

The modulus masses are given by \cite{Conlon-Quevedo} 
\beq
m \sim \frac{ln\mathcal{V}}{\mathcal{V}} m_{p}.
\eeq
It is useful to parametrize $m$ in terms of $\alpha$. Since we are 
interested in $\alpha \sim 3\times 10^{7}$, the factor $ln\mathcal{V}$ can
be  replaced by $35$. Hence, we obtain for the mass
\beq\label{LVmass}
m \sim 35 m_{p} \alpha^{-2}.
\eeq
For $\alpha \sim 3\times 10^{7}$  we have $m \sim 4 \times 10^{5}$ GeV,  
and we fix $\mu_{-20} \sim 10$, to ensure $z_* > z_\nu$.  
The last condition means choosing the 
basic cosmic string parameter, the symmetry breaking string scale 
$\eta = \sqrt{\mu} \sim 3.9 \times 10^9$~ GeV, i.e. 
$G\mu \sim 1\times 10^{-19}$.  With this choice of parameters, 
$z_*$ from Eq.~(\ref{z*}) is given by $z_* \approx 250$, i.e it is
larger than $z_{cas}^{\max} \sim 60$ and than the neutrino 
horizon $z_\nu \sim 200$. For this case, the neutrino flux is given by 
Eq.~(\ref{E2Jnunew}) and the cascade energy density by Eq.~(\ref{cas-density1}).
With the parameters $\alpha_7$ and $m_5$ as indicated above, the
neutrino flux is $E^2J(E) = 1.3 \times 10^{-8} p^{-1}$~ GeV/cm$^2$~s~sr and
the cascade energy density is 
$\omega_{cas} \approx 2.6\times 10^{-8} p^{-1}$~ eV/cm$^3$.  

This neutrino flux is shown in Fig.~\ref{fig1} by the curve ``theor.2". It is
detectable by LOFAR and SKA, but not by JEM-EUSO.  

We can modify this model choosing parameters providing a larger
neutrino flux. For this we fix $\alpha_7=2$,~ $m_5=0.1$ and keep
$z_*=250$, i.e. $\mu_{-20} \approx 80$ and $\eta \approx 1\times
10^{10}$~GeV. The cascade energy density, 
$\omega_{\rm cas}=7.3\times 10^{-8} p^{-1}$~eV/cm$^3$, is safely below 
$\omega_{\rm cas}^{\max}$.  This model satisfies the restrictions 
obtained in \cite{Sabancilar}, and in fact we can further increase the
flux by decreasing $m_5$. At high energy limit the calculated neutrino flux is given by 
$E^2 J_\nu(E) \approx 3.2\times 10^{-8} p^{-1}$~ GeV/cm$^2$ s sr.
This flux is shown in Fig.~\ref{fig1} 
by curve theor.1, it is detectable by JEM-EUSO, LOFAR and SKA.

\subsection{An illustrative example for low $z_*$ model with $z_* < z_\nu$}
\label{example2}

Let us now assume that $z_* < z_\nu$.  Then, 
in the redshift interval $z_* \leq z \leq z_\nu$ the modulus 
radiation dominates. In the remaining interval 
$0 \leq z \leq z_*$ the 
gravitational energy losses are dominant. The total neutrino 
flux is given by the sum of fluxes generated from both intervals.
A value of $z_*$ fixes a model. Here we try to find a model with a
detectable neutrino flux. 

The neutrino flux from the moduli-dominated interval $(z_*\; , z_\nu)$,
in the HE asymptotic regime, is given by Eq.~(\ref{E2JnuHE}). This equation 
becomes valid just above the low-energy steepening, i.e. at 
\beq 
E \geq \tilde{E}_{\min} = 1\times 10^{19}\; \alpha_7^2\; (50/z_*)^2~{\rm eV}.
\label{tildeE}
\eeq

For the  flux from the gravity-dominated interval $(0,\; z_*)$, the HE 
asymptotic is given 
by Eq.~(\ref{E2Jnu4HE}), with a low-energy cutoff approximately at the
same energy as above. Summing up both components, we obtain  
\beq
E^2 J_\nu (E)= 1.8 \times 10^{-10} p^{-1} \alpha_7^2 m_5^{-1/2} 
(1+z_*)^{1/2}\left [2- \left (\frac{1+z_*}{1+z_\nu}\right )^{1/2}\right ]
~ {\rm GeV~ cm}^{-2}~{\rm s}^{-1}~ {\rm sr}^{-1}.
\label{E2JnuHE-sum}
\eeq

To maximize the flux without strongly increasing the low-energy
steepening threshold  $\tilde{E}_{\min}$ given by  Eq.~(\ref{tildeE}), one can use 
the parameters: $\alpha_7=2,\; m_5=0.1$  and $z_* = 100$. As a result 
we obtain $\tilde{E}_{\min} = 1\times 10^{19}$~ eV and 
$E^2J_{\nu}=2.9 \times 10^{-8}$~ GeV~ cm$^{-2}$~ s$^{-1}$~ sr$^{-1}$,
close to the upper limit ``$E^{-2}$-cascade" in Fig.~\ref{fig1}.
This flux is detectable by JEM-EUSO, LOFAR and SKA. The cascade energy
density does not exceed $\omega_{\rm cas}^{\max}$.

For cosmic F-and D-strings, the $p^{-1}$ factor alone can increase the flux
and $\omega_{cas}$  to and above ``E$^{-2}$ cascade limit", keeping 
$E_{\min}$ unchanged.

\subsection{EHE neutrino detection}
\label{NeutrinoDetection}

The calculated EHE neutrino fluxes are shown in Fig.~\ref{fig1},
together with the existing upper limits from ANITA 08 \cite{anita}, 
ANITA-II \cite{anita}, 
RICE\cite{rice}, and with the sensitivity of the proposed experiments -- 
JEM-EUSO \cite{JEM} 
(to be launched in 2015), LOFAR \cite{lofar} and SKA \cite{ska}. The 
predicted fluxes are presented for 
gravity-dominated regime with $z_* > z_\nu$. As is discussed 
in Section \ref{example2} in moduli dominated regime the neutrino
fluxes can be also detectable. The characteristic feature of all
models is very high minimal energy of neutrinos in spectrum $E_{\min}$.
Because of this, neutrino fluxes in some models are undetectable by JEM-EUSO. 

In all cases neutrino fluxes and cascade energy density are scaled 
by factor $p^{-1}\,\alpha_7^2\;m_5^{-1/2}$. The quantity
$\alpha_7^2\;m_5^{-1/2}$ alone can be increased by factor 10-30. 
In our models neutrinos are produced at large redshifts, while 
$\omega_{\rm cas}^{\max}=5.8\times 10^{-7}$~eV/cm$^3$ is obtained 
mainly due to observation of 100~GeV photons \cite{BGKO}, which have a
local origin. Increasing the value of $\alpha_7$ is limited since 
$E_{\min}$ is usually proportional to $\alpha_7^2$ and these high 
energy neutrinos become undetectable by JEM-EUSO.  The quantity $p^{-1}$
can be easily increased by factor $10^3$ for cosmic  F- and D-strings.
This factor is limited by $\omega_{\rm cas}$ only, $E_{\min}$ does not 
depend on $p^{-1}$.

\begin{figure}[htb]
\centering\includegraphics[height=8cm,width=12cm]{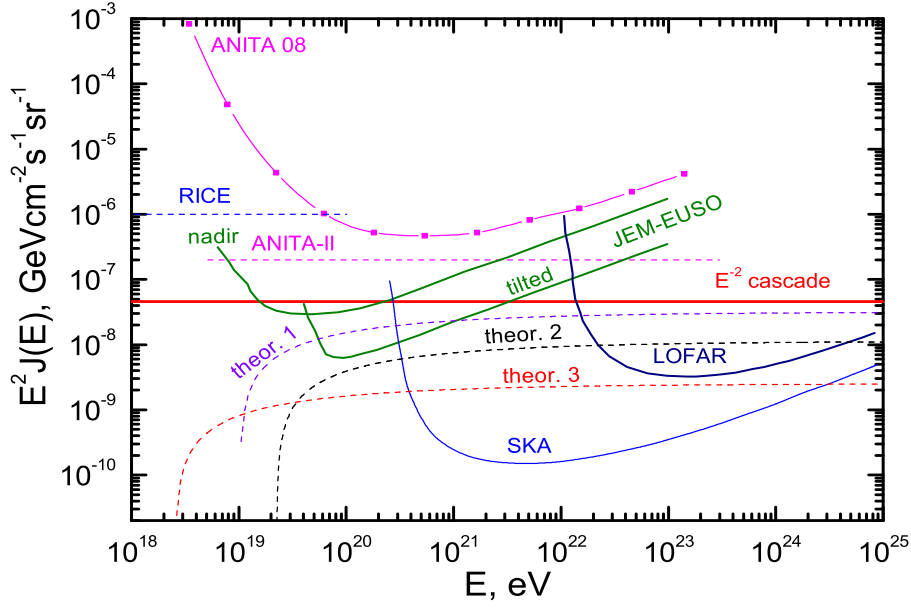}   
\caption{{\em Calculated neutrino fluxes compared with existing upper
  limits, ANITA-II, RICE, ANITA 08, and with sensitivities of
  projects, JEM-EUSO, LOFAR and SKA.} Line ``E$^{-2}$ cascade" presents 
  the upper limit for {\em cosmogenic} neutrinos \cite{BGKO}. Curve 
  ``theor.3" gives the predicted neutrino flux for gravity-dominated 
  regime with ``normalizing" set of parameters  $p=1$ (ordinary
  strings), and $\alpha_7=m_5=z_*/z_\nu =1$. This flux is detectable 
  only by SKA. Curve ``theor.2" corresponds to superstring-motivated
  example in gravity-dominated regime (see Section \ref{example1})
  with $p=1$,~ $z_*/z_\nu=1$,~ $\alpha_7=3$ and $m_5=4$. This flux is
  detectable by SKA and LOFAR. Increasing further the product 
  $\alpha_7^2 m_5^{-1/2}$ the flux can be made detectable by JEM-EUSO,
  though increasing of $\alpha_7$ should be limited, because of increasing
  of $E_{\min}$ due to this factor. The curve ``theor.1" with 
  parameters $p=z_*/z_\nu=1$,~ $\alpha_7=2$ and $m_5=0.1$ gives the
  flux detectable by all three future instruments SKA, LOFAR and JEM-EUSO.
  In case of moduli-dominated regime with parameters considered in 
  Section \ref{example2} as $\alpha_7=2$,~ $m_5=0.1$, and $z_*=100$ 
  neutrino flux is observable by all three detectors, SKA, LOFAR and 
  JEM-EUSO.  
}
\label{fig1}
\end{figure}

The characteristic feature of our model
is the production of EHE neutrinos due to
tremendously large Lorentz factors. The neutrino-induced gigantic 
showers in the air or lunar regolith produce a strong signal 
reliably detectable in optical and radio emissions. The signature 
of EHE neutrinos with energies above $10^{21} - 10^{22}$~ eV is 
given by this energy scale, which is unaccessible for astrophysical sources. 

This model has another signature, already 
discussed in \cite{BOSV}: the simultaneous appearance of a few  
showers in the field of view of a detector. It is due to neutrino 
propagation in the form of a very narrow jet
and to the absence of time-delay in neutrino arrival, because of the   
tremendously large neutrino Lorentz factors $\Gamma_\nu = E_\nu/m_\nu$. 
Compared to the case of superconducting strings 
in \cite{BOSV}, the rate of multiple showers is strongly suppressed by 
higher energies of neutrinos, being partly compensated by  
the greater target mass of gigantic radio detectors. We present 
some brief estimates below.  

The number of detected neutrinos from a jet in the detector target 
is given by   
\beq 
N_\nu^{det}(>E) = \frac{\sigma_{\nu N}(>E)}{m_N} M_{det} F_\nu(>E),
\label{Nnu}
\eeq
where $\sigma_{\nu N}(>E)$ is the
neutrino-nucleon cross-section for neutrinos with energies greater 
than $E$, $M_{det}$ is the target mass for the largest 
radio-detectors, and $F_\nu(>E)$ is the fluence of neutrinos with energy 
greater than $E$ in a jet from a source at redshift $z$. 
The fluence is calculated as 
\beq 
F_\nu(>E)= \int dk N_X^b (k) \frac{\xi_\nu(k,z, >E_{\min})}{\Omega_k r^2} ,   
\label{fluence-gen}
\eeq
where $E_{\min}$ is the minimum neutrino energy for a source at 
redshift $z$.  In numerical estimates we shall use 
$\sigma_{\nu N} \sim 1\times 10^{-31}$~cm$^2$ and 
$M_{det} \sim 10^{21}$~g . 

For the gravity-dominated regime, standard calculations give the 
following expression for the neutrino fluence from a loop of length 
$L_{\min}$ at redshift $z$:
\beq
F_\nu (>E) = \frac {\alpha^2}{3\pi 2^{5/3}}\frac{f_\pi}{b_*} 
\frac{\Gamma^{3/2} (G\mu)^{7/2}}{(t_0 m)^{1/2}}
\frac{1}{[(1+z)^{1/2} - 1]^2(1+z)^{9/4}} \frac{m}{E_{\min}} m_{Pl}^2 ,
\label{fluence-grav-regime}
\eeq
where for $z < 1$~ 
$E_{\min} = 4.5\times 10^{22} m_5^{1/2}\mu_{-20} \varepsilon$~eV . 
Estimating fluence for the illustrative case parameters $\alpha=3\times 10^7$,
$m=4 \times 10^5$~ GeV, $\mu_{-20}=10$, and using $z \sim 1$ and 
$E_{\min} \sim 10^{13}$~GeV, we obtain 
$F_\nu \sim 3\times 10^{-18}$~ cm$^{-2}$. The number of neutrinos
detected per burst is calculated as   
$N_\nu^{det} \sim 2\times 10^{-4}\mu_{-19}^{3}$.  The fact that 
$N_\nu^{det} \ll 1$ shows that practically all detected neutrinos are single.
This is mostly due to the large value of $E_{\min}$ in the 
gravity-dominated regime. 

The situation is different for the moduli-dominated regime at 
$z_* \leq z \leq z_{\nu}$. The fluence 
from a loop with $L\sim L_{\min}$ at redshift $z_*$ is given by 
\beq
F_\nu(>E)=\frac{2^{-2/3}}{6\pi}\frac{f_\pi}{b_*}(G\mu)^3 \frac{1}
{(1+z_*)^{3/2}[(1+z_*)^{1/2} - 1]^2} \alpha^4 \frac{m_{Pl}^2}{E t_0}.
\label{fluence-moduli-regime}
\eeq
For the model with $\alpha_7=10$, $m_5=1$ and $z_*=40$, which yields 
the e-m cascade flux at the energy density bound 
$\omega_{cas}^{\max}$,
and assuming the neutrino energy $E=1\times 10^{20}$~eV, we obtain numerically 
$F_\nu(>E)=6.2\times 10^{-15}$~cm$^{-2}$, and the mean number of 
neutrinos detected in a burst is $\bar{N}_\nu^{det} \sim 0.4$.  The Poisson 
probability to detect simultaneously  $n=2$ neutrinos at average 
$\bar{N}=0.4$ is 0.054.


\section{Conclusions}

Production of high energy particles is a natural feature and one of
the signatures of topological defects,  including cosmic strings.  
It provides a
method of searching for, e.g., cosmic strings, which complements other methods, 
based on the gravitational effects of strings, such as structure formation, 
CMB data, gravitational radiation, 
gravitational lensing and others. The strongest current bounds on
strings with a symmetry breaking energy scale $\eta=\sqrt{\mu}$ is given 
by $G\mu \lsim 10^{-7}$ due to lensing effect \cite{lensing} and 
$G\mu \lesssim 4\times 10^{-9}$, due to the millisecond pulsar observations 
\cite{pulsar}.  With more advanced 
gravitational wave detectors, the bound is expected to improve to 
$G\mu \sim 10^{-12}$ \cite{Damour05,grav-waves}.  On the other hand, UHE
particles can be detected from strings with $G\mu$ values as small as
$\sim 10^{-20}$. 

Cosmic strings can arise from a symmetry breaking phase transition in 
the early universe. Fundamental strings of superstring theory can also 
play the role of cosmic strings in some models.  

A characteristic feature  of the string dynamics is the periodic 
appearance of cusps, where very 
large Lorentz factors are reached for brief periods of time.  Particle 
emission from cusp segment results in extremely
high particle energies. Astrophysical sources, at the present level of 
knowledge, 
cannot accelerate particles to energies above $10^{21} - 10^{22}$~eV, 
with the maximum neutrino energy an order of magnitude lower. Thus,  
detection of neutrinos with $E \gsim 10^{21}$~eV would be a signature of 
top-down models, with the string-cusp model as a plausible candidate. 
An additional signature of this model is neutrino emission in the form 
of a narrow jet with simultaneous detection of two or more neutrinos 
possible in the field of view of the detector. However, this
possibility exists 
only for some values of model parameters and only for large detectors.    

EHE neutrino astronomy with $E \gsim 10^{21}$~eV can probe 
high energy processes in the universe up to red-shifts 
$z_\nu \sim 200$. At the same time these neutrinos have large 
interaction cross-section with nucleons, 
$\sigma_{\nu N} \sim 1\times 10^{-31}$~cm$^2$, favorable for detection. 

Moduli are produced near cusps of oscillating string loops,
where the characteristic frequency of string motion exceeds the 
modulus mass $m$. 
These particles are emitted from a string segment with a Lorentz factor 
$\gamma_c \sim \sqrt{mL}/4$, where $L$ is the length of the loop.
Moduli and the products of their decays move as a jet with an opening 
angle $\theta_c \sim \gamma_c^{-1}$. A modulus decays into two gluons, 
which initiate a quark-gluon cascade, which turns into hadrons, mostly 
pions, and then to neutrinos. We assume that the neutrino spectrum in the
rest frame of the modulus is $\propto E_*^{-2}$, where $E_*$ is 
the neutrino energy in this frame. We adopted this spectrum in order 
to simplify the analysis and to obtain  transparent analytic results. In 
fact the neutrino spectrum is not power-law, and has a flattening at a 
low energy $\epsilon_*^{\min}$, which we call the minimum energy.  
This spectrum is Lorentz boosted, as described in Section \ref{boost},
and the boosted spectrum also has a flattening at energy 
$E_{min} \sim \gamma_c \epsilon_*^{\min}$.  This energy can be
considered as an effective low-energy cutoff of the spectrum, because  
at $E < E_{\min}$ the detectability of neutrinos is noticeably reduced.

Given the length distribution of loops at epoch $z$ and the spectrum of moduli emitted from a loop of a given length, it is possible to calculate the neutrino flux
$E^2 J_\nu (E)$, the cascade energy density $\omega_{cas}$, which 
provides an upper limit on the neutrino flux (see Section 
\ref{CascadeBound}), and the cutoff energy $E_{\min}$ in the neutrino spectrum. 
The results are given in terms of three free parameters 
$\alpha_7=\alpha/10^7$,~ $m_5=m/10^5$~GeV and 
$\mu_{-20}=G\mu/10^{-20}$. Another important parameter is 
the redshift of the neutrino horizon, $z_\nu \sim 200$.

The results of these calculations critically depend on the redshift 
$z_* \approx 400 \alpha_7^{-8/3} m_5^{2/3} \mu_{-20}^{2/3}$. 
This redshift separates two regimes in the string evolution:
the gravity-dominated regime at $z \leq z_*$, when gravitational energy 
losses are dominant, and the moduli-dominated regime at $z \geq z_*$, 
when modulus radiation energy losses dominate. 

Apart from this selective role, fixing the value of $z_*$ gives a  
constraint in the parameter space ($\alpha_7$,~ $m_5$,~ $\mu_{-20}$),  
in the form $z_*(\alpha_7, m_5, \mu_{-20}) = z_*'$, where $z_*'$ is 
the fixed $z_*$ value. This reduces the number of free parameters to
two (at the fixed $z_*$), which we choose as  $\alpha_7$ and $m_5$.
As a result all calculated fluxes $E^2 J_\nu (E)$ and energy density 
$\omega_{cas}$ scale (at fixed $z_*$) as $p^{-1} \alpha_7^2 m_5^{-1/2}$.

In our calculations the fixed value of $z_*$ plays the role of the most
important parameter, which determines what we call ``the model".
The high $z_*$ models, defined as $z_* > z_\nu$ correspond to the
gravity-dominated regime in the whole allowed interval of redshifts 
$(0,\; z_\nu)$. The low $z_*$ models, defined as $z_* < z_\nu$, are
characterised by two regimes: the gravity dominated one  at
$0 \leq z \leq z_*$ and the moduli-dominated one at 
$z_* \leq z \leq z_\nu$, with approximately equal neutrino fluxes. 
The total neutrino flux is increasing with
growth of $z_*$ until it reaches $z_\nu$, and models with 
$z_* \geq z_\nu$ give the largest flux. This growth of flux with $z_*$ is 
accompanied by a decrease of $E_{\min}$, 
see Eq.~(\ref{Emin-both}), which is favorable  for neutrino detection
by JEM-EUSO. 


Three theoretical curves in Fig.~\ref{fig1} illustrate different cases 
of neutrino detectability for the gravity-dominated regime at 
$z_* \geq z_\nu$. The predicted  flux is detectable by SKA only  in
case of ``normalizing" parameter set $p=\alpha_7=m_5=z_*/z_\nu=1$
(curve ``theor.3"). The flux is detectable by LOFAR and SKA 
in case of $p=z_*/z_\nu=1$,~ $\alpha_7=3$~ and $m_5=4$ (curve 
``theor.2"). The flux is delectable by all three detectors JEM-EUSO, 
LOFAR and SKA if  $p=z_*/z_\nu=1$,~ $\alpha_7=2$~ and $m_5=0.1$  
(curve ``theor.1").

We considered above ``ordinary" strings with $p = 1$. For cosmic
superstrings with reconnection probability $p< 1$, the neutrino flux
increases by a factor $p^{-1}$ without increasing $E_{min}$, and is detectable for a wider range of model parameters.

The cascade upper limit on neutrino flux in Fig.~\ref{fig1} 
($E^{-2}$ curve) is given for cosmogenic neutrinos from \cite{BGKO}. 
It is based on maximum energy density $\omega_{\rm cas}^{\max}$
allowed by Fermi data \cite{Fermi}. For the the considered model with 
the dominant contribution from large redshifts this upper limit is
higher because of 100~GeV-neutrino absorption. Therefore, allowed
neutrino fluxes can be further increased. 

The remarkable feature of moduli-produced strings is strong beaming. 
The Lorentz-factor at $z=z_\nu$ in gravity-dominated regime and at 
$z_*$ in moduli-dominated regime is $\Gamma \sim 10^{12}$. The
corresponding angle of a beam is 
$\theta \sim \Gamma ^{-1} \sim 10^{-12}$. Neutrinos with this
tremendous energies arrive simultaneously at a detector and can
produce simultaneously two or more showers in the field of view of detector. 
The estimates made in Section \ref{NeutrinoDetection} shows that such 
events are rare due to very high energies of neutrinos.


\section*{ACKNOWLEDGMENTS}
The authors are grateful to Peter Gorham for valuable correspondence
and to Askhat Gazizov for preparing the figure
and for useful discussions. The work by E.S. and A.V.
was supported in part by the National Science Foundation under grant PHY-0855447.


\section*{APPENDIX}

In this section we shall derive the modulus radiation spectrum in more detail. In a flat background, i.e., $g_{\mu\nu} = \eta_{\mu\nu} = diag(-1, 1, 1,1)$, the equation of motion for the string worldsheet $X^\mu(\sigma^a)$ is
\beq
\partial_{a} \left(\sqrt{-\gamma} \gamma^{ab} X^{\mu}_{,b}\right) = 0.
\eeq
Using the conformal gauge and $\sigma^{0} \equiv \tau$, $\sigma^{1} \equiv \sigma$ one obtains
\beq\label{stringEOM} 
\ddot{X}^{\mu} - {X^{\prime \prime}}^{\mu} = 0,
\eeq
and the gauge conditions are
\beq
\dot{{\bf X}}{\bf \cdot X}^{\prime} = 0 ,
\eeq
\beq
\dot{{\bf X}}^{2} + {{\bf X}^{\prime}}^{2} = 1.
\eeq
In this gauge, the worldsheet coordinate $\tau$ can be identified with the Minkowski time coordinate $t$. The solution for (\ref{stringEOM}) can be written in terms of the right moving and the left moving waves as
\beq
{\bf X}(\sigma, \tau) = \frac{1}{2}\left[{\bf X}_{+}(\sigma_{+}) + {\bf X}_{-}(\sigma_{-}) \right] ,
\eeq
where the lightcone coordinates are defined as $\sigma_{+} \equiv \sigma + \tau$, $\sigma_{-} \equiv \sigma - \tau$. The corresponding gauge conditions are ${{\bf X}_{+}^{\prime}}^{2} = {{\bf X}_{-}^{\prime}}^2 = 1$, where primes denote derivatives with respect to the lightcone coordinates. 

Using the lightcone coordinates, Eq.(\ref{Tk}) can be written in the form
\beq\label{energy-momentum-fourier}
T({\bf k}, \omega_{n}) = - \frac{\mu}{L} \int_{-L}^{L} d\sigma_{+}
\int_{-L}^{L} d \sigma_{-} \left(1+
{\bf X}_{+}^{\prime}{\bf \cdot X}_{-}^{\prime}\right)
e^{\frac{i}{2}\left[(\omega_{n} \sigma_{+} - {\bf k\cdot X}_{+}) -
(\omega_{n} \sigma_{-} + {\bf k \cdot X}_{-}) \right]}.
\eeq

Since we shall be mainly interested in modulus bursts from
cusps, we use the expansion of string worldsheet about a cusp, which
we take to be at $\sigma_+ =\sigma_- =0$. The functions in the
integrand of (\ref{energy-momentum-fourier}) can be calculated from
the expansions as
\beq
1+ {\bf X}_{+}^{\prime}{\bf \cdot X}_{-}^{\prime} \approx - \frac{4\pi^{2} s}{L^2} \sigma_{+} \sigma_{-},
\eeq
and
\beq
{\bf k \cdot X}_{\pm} \approx k \left(\pm \sigma_{\pm} \mp \frac{2 \pi^2}{3 L^{2}}\sigma_{\pm}^3\right),
\eeq
where $s$ is an $O(1)$ parameter which depends on the loop trajectory and ${\bf k}$ is assumed to be in the direction of the string
velocity at the cusp.  

Eq. (\ref{energy-momentum-fourier}) can now be separated into two
integrals as
\beq\label{energy-momentum-cusp-fourier}
T({\bf k}, \omega_{n}) = \frac{4 \pi^2 \mu\, s}{L^{3}} I_{+} I_{-} ,
\eeq
where
\beq\label{integral-pm}
I_{\pm} = \int_{-L}^{L} d\sigma_{\pm}\, \sigma_{\pm}\, e^{\pm i
\left[\frac{\omega_{n}-k}{2}\sigma_{\pm} + \frac{\pi^2 k}{3 L^2}
\sigma_{\pm}^{3}\right]}.
\eeq
After a change of variables, we obtain the integral
\beq\label{integral-pm-u}
I_\pm (u) = \frac{L^2}{2 \pi^2} \left(\frac{\omega_{n}}{k} -1\right)
\int_{- \infty}^{\infty} dx\, x\, e^{\pm i \frac{3}{2}u \left[x +
\frac{1}{3} x^{3}\right]} ,
\eeq
where
\beq\label{u}
u \equiv \frac{L k}{3 \sqrt{2} \pi} \left(\frac{\omega_{n}}{k}-1\right)^{3/2},
\eeq
and we have approximated the upper and lower limits of integration as
$\pm \infty$.  
The real part of the integral is zero since
it is an odd function of $x$. The imaginary part can be expressed in
terms of the modified Bessel function,
\beq
I_\pm (u) =\pm i \frac{L^2}{2 \pi^2} \left(\frac{\omega_{n}}{k} -1\right)
\frac{2}{\sqrt{3}} K_{2/3}(u).
\eeq
Then, (\ref{energy-momentum-cusp-fourier}) can be calculated as
\beq
T({\bf k}, \omega_{n}) = \frac{4 L \mu \, s}{3 \pi^2}
\left(\frac{\omega_{n}}{k} -1\right)^{2} K_{2/3}^{2}(u),
\eeq
and the power spectrum for the moduli radiation
(\ref{power-spectrum}) from a cusp is
\beq\label{power-spectrum-cusp}
\frac{d P_{n}}{d \Omega} = \frac{8 L^2 \alpha^2 s^{2} G \mu^2}{9 \pi^5} 
\omega_{n} k \left(\frac{\omega_{n}}{k} -1\right)^{4} K_{2/3}^{4}(u).
\eeq

The asymptotic form of the power spectrum for $k>>m$ and $\omega_{n}
\approx k$, i.e., $u<<1$, is\footnote{When $u<<1$, $K_{\nu}(u) \approx
\frac{\Gamma(\nu)}{2} \left(\frac{2}{u}\right)^{\nu}$.}
\beq
\frac{d P_{n}}{d \Omega} \approx \tilde{\Gamma} \alpha^2 s^2 
G \mu^2 n^{-2/3}.
\eeq
where $\tilde{\Gamma} \sim 1$. This is the same as the power spectrum for gravitons, except that for
gravitons there is no additional coupling constant $\alpha$ and the numerical coefficient is somewhat different.

The average rate of moduli radiation per solid angle is
\beq\label{moduli-rate-solid-angle}
\frac{d \dot{N}}{d \Omega} = \sum_{n} \frac{1}{\omega_{n}} 
\frac{d P_{n}}{d \Omega}.
\eeq
The sum over $n$ can be converted into an integral over $k$ by using
the relation $\omega_{n} = \frac{4\pi n}{L} = \sqrt{k^2 +m^2}$
\beq\label{sum-n}
\sum_{n} = \frac{L}{4 \pi} \int \frac{k\, d k}{\sqrt{k^2 + m^2}}. 
\eeq
Here we only consider the modulus bursts which have very large Lorentz
factors, thus we keep the leading order term in the limit $k>>m$. In
this limit, (\ref{u}) becomes $u \approx \frac{L m^3}{12 \pi k^2}$. By
substituting (\ref{power-spectrum-cusp}) into
(\ref{moduli-rate-solid-angle}), using (\ref{sum-n}) and also by
making a change of variable $u \equiv \frac{L m^3}{12 \pi k^{2}}$, we
obtain
\beq\label{rate-integral}
d \dot{N} \sim \frac{\alpha^2 s^2 G 
\mu^2}{m}  K_{2/3}^{4} (u) u^2 du d\Omega .
\eeq

The function $K_{2/3}(u)$ dies out exponentially at $u \gtrsim
1$. Hence, the main contribution to the rate comes from the region $u
\lesssim 1$ which corresponds to
\beq
k\gtrsim k_{min} = k_{c} \sim \frac{1}{4} m \sqrt{mL}. 
\label{kmin}
\eeq
For $k>> k_{min}$, Eq.~(\ref{rate-integral}) gives
\beq\label{dNdot}
d \dot{N} \sim \alpha^2 s^2 G 
\mu^2 L^{1/3} k^{-5/3} dk d\Omega .
\eeq
From (\ref{dNdot}), the number of moduli emitted from a cusp in a
single burst, into solid angle $d\Omega$, having momentum between
$(k,k +dk)$ can be estimated as
\beq\label{dN}
dN \sim L d\dot{N} \sim \alpha^2 s^2 G \mu^2 L^{4/3} k^{-5/3} dk d\Omega.  
\eeq
Here, we assumed one cusp event per oscillation period of a
loop.

Moduli are emitted into a narrow opening angle around the
direction of the string velocity ${\bf v}$ at the cusp.  The spectral
expansion (\ref{dNdot}) has been calculated for moduli emitted in
the direction of ${\bf v}$.  For moduli emitted at a small angle
$\theta$ relative to ${\bf v}$, Eq.~(\ref{dNdot}) still applies, but
now the spectrum is cut off at $k_{max}\sim 1/L\theta^3$.  In other
words, the opening angle for the emission of particles with momenta
$\gtrsim k$ is
\beq\label{theta_k}
\theta_k \sim (kL)^{-1/3}.
\eeq
Integration over $\Omega$ in (\ref{dNdot}) gives a factor $\sim
\theta_k^2$,
\beq\label{dNdks}
d \dot{N} \sim \alpha^2 s^2 G 
\mu^2 L^{-1/3} k^{-7/3} dk .
\eeq
The dominant contribution to the modulus emission comes from $k\sim
k_{min}$, and the total emission rate is
\beq
\label{total-moduli-rate}
\dot{N} \sim \alpha^2 s^2 G 
\mu^2 L^{-1/3} k_{min}^{-4/3}.
\eeq
The corresponding opening angle is
\beq
\theta_{c}\sim \gamma^{-1}\sim m/k_{min}.
\eeq
The total power of the modulus radiation can be similarly calculated
as
\beq
P\sim \alpha^2 s^2 G \mu^2 L^{-1/3} k_{min}^{-1/3}.
\eeq


\end{document}